\documentclass[sigconf]{acmart}

\usepackage{algorithm}
\usepackage{algpseudocode}
\AtBeginDocument{%
  \providecommand\BibTeX{{%
    \normalfont B\kern-0.5em{\scshape i\kern-0.25em b}\kern-0.8em\TeX}}}

\setcopyright{acmlicensed}
\copyrightyear{2018}
\acmYear{2018}
\acmDOI{XXXXXXX.XXXXXXX}

\acmConference[Conference acronym 'XX]{Make sure to enter the correct
  conference title from your rights confirmation emai}{June 03--05,
  2018}{Woodstock, NY}
\acmBooktitle{Woodstock '18: ACM Symposium on Neural Gaze Detection,
 June 03--05, 2018, Woodstock, NY} 
\acmISBN{978-1-4503-XXXX-X/18/06}

\usepackage{amsmath}
\usepackage{graphicx} 
\usepackage{subcaption} 

\begin{document}

\title{Offensive AI: Enhancing Directory Brute-forcing Attack with the Use of Language Models}




\author{Alberto Castagnaro}
\affiliation{%
 \institution{Delft University of Technology} 
 \country{The Netherlands}}
\email{A.Castagnaro@student.tudelft.nl}

\author{Mauro Conti}
\orcid{0000-0002-3612-1934}
\affiliation{%
 \department{Department of Mathematics}
 \institution{University of Padova}
 \city{Padova}
 \country{Italy}}
  \additionalaffiliation{%
 \institution{Delft University of Technology}
 \country{The Netherlands}}
\email{mauro.conti@unipd.it}

\author{Luca Pajola}
\orcid{0000-0002-6749-6608}
\affiliation{%
 \institution{Spritz Matter Srl} 
 \city{Padova}
 \country{Italy}}
 \additionalaffiliation{%
 \institution{University of Padova}
 \city{Padova}
 \country{Italy}}
\email{luca.pajola@unipd.it}






\renewcommand{\shortauthors}{Trovato and Tobin, et al.}

\begin{abstract}
Web Vulnerability Assessment and Penetration Testing (Web VAPT) is a comprehensive cybersecurity process that uncovers a range of vulnerabilities which, if exploited, could compromise the integrity of web applications. 
In a VAPT, it is common to perform a \textit{Directory brute-forcing Attack}, aiming at the identification of accessible directories of a target website. 
Current commercial solutions are inefficient as they are based on brute-forcing strategies that use wordlists, resulting in enormous quantities of trials for a small amount of success. 
\par
Offensive AI is a recent paradigm that integrates AI-based technologies in cyber attacks. In this work, we explore whether AI can enhance the directory enumeration process and propose a novel Language Model-based framework. 
Our experiments -- conducted in a testbed consisting of 1 million URLs from different web application domains (universities, hospitals, government, companies) -- demonstrate the superiority of the LM-based attack, with an
average performance increase of 969\%.  
\end{abstract}




\maketitle

\section{Introduction}

In its most general sense, hacking refers to modifying or manipulating a system's features to achieve a goal outside of the creator's original purpose. While often associated with illegal cyber activities, hacking can also be performed ethically, with permission, to improve system security and uncover vulnerabilities that can be fixed before malicious actors exploit them.
\par
Directory enumeration is a critical component of security assessments. It involves identifying accessible directories, files, and web paths in a web application. Effective discovery attacks, such as directory brute-forcing, can uncover hidden directories and files that may contain sensitive data or critical functionalities.
\par
Offensive AI uses artificial intelligence technologies to conduct or enhance cyber attacks~\cite{kaloudi2020ai, mirsky2023threat}. This emerging field combines AI's adaptability and learning capabilities with traditional attack vectors, creating more sophisticated and automated threats. Offensive AI can rapidly analyze vast amounts of data, adapt to defensive measures, and execute attacks with increased speed and complexity.
\paragraph{Contribution}
This paper contributes to the field by presenting a novel approach that leverages Language Models (LMs) to enhance the efficiency and effectiveness of directory brute-forcing attacks. Our method builds on prior knowledge retrieved by different web applications and then exploits embedding to extrapolate the context of different words that form web paths and language models to generate new possible URL (Uniform Resource Locator) paths that can be used to send requests.
Our contributions are summarized below:
\begin{enumerate}
    \item We designed a novel dataset containing $4$ distinct types of applications that are often the targets for attacks, i.e., commercial, government, hospital and universities, for a total of $1$ million of URLs. 
    \item We propose two novel directory brute-forcing attacks that leverage prior knowledge: a probabilistic and a Language Model-based approach.
    \item A systematic evaluation highlights the superiority of prior knowledge approaches compared to baselines. LM-based attacks outperform all 8 proposed baselines, with an average performance increase of 969\%. On the other hand, probabilistic approaches show high performance when the budget of spendable requests is limited and, therefore, optimal for stealthier attacks.   
\end{enumerate}

\paragraph{\textbf{Ethical Disclaimer}} The techniques and methods discussed in this paper are intended for educational purposes and ethical security testing only. The authors do not condone the use of these methods for malicious purposes and strongly advocate for responsible disclosure and remediation of identified vulnerabilities. We hope that this research will contribute to the development of more secure web environments and the advancement of cybersecurity practices.
For this reason, we do not share publicly the collected dataset and code. Researchers willing to reproduce our experiment are invited to contact the authors. 

\section{Background}\label{sec.background}
This section describes the theory behind Language Models~\cite{pennington2014glove, schnabel2015evaluation, petroni-etal-2019-language}. 
Language Models (LMs) are statistical models that learn the \textit{probability distribution} of sequences of words in a language. Their objective is to predict the likelihood of a word given the context of preceding words. Formally, given a sequence of words $\mathbf{x} = (x^{(1)}, \ldots, x^{(t)})$, LMs computes the probability distribution of the next word $x^{(t + 1)}$:
\begin{equation}
    P(x^{(t+1)} | x^{(t)}, \ldots x^{(1)}),
\end{equation}
where $x^{(t + 1)} \in V = {w_1, \ldots, w_{|V|}}$, and $V$ is a fixed vocabulary. 
Given a sentence, the goal of a LM is to estimate the probability of this sequence $P(\mathbf{x})$, which is obtained through the chain rule of probability: 
\begin{equation}
\begin{aligned}
        P((x^{(1)}, \ldots, x^{(T)})) & = P(x^{(1)}) \cdot p(x^{(2)} | x^{(1)}) \cdot \ldots \cdot \\
        & p(x^{(T)} | x^{(T - 1)}, \ldots, x^{(1)}) \\
        & = \prod_{t=1}^{T}(x^{(t)}|x^{(t-1)}, \ldots, x^{(1)}).
\end{aligned}
\end{equation}

Modern LMs utilize neural networks to learn complex relationships between words and context.
Recurrent Neural Networks (RNN) are ideal for the task, as they model the generative process of a sequence data (\textit{e.g.,} time series, natural language). 
Unlike other types of NN like feedforward NN, RNNs integrate feedback connections that allow them to retain information from previous time steps (\textit{hidden state}), and then generate a new sample according to a probability distribution given the hidden state. 
At each time step, a RNN computes an output $y^t$ based on the current input $x^t$ and the hidden state $h^{t-1}$ calculated at the previous step. 
\begin{equation}
    h^{(t)} = f(h^{(t - 1)}, x^{(t)}). 
\end{equation} 
LMs are trained on large text corpora, and their parameters are learned to maximize the likelihood of observed sequences (maximum likelihood estimation). 
The loss function at step $t$ is the cross-entropy between predicted probability distribution $\hat{y}_t$ and the true next word $y_t$:
\begin{equation}
    \begin{aligned}
            J^{(t)}(\theta) &  = CE(\mathbf{y}^{(t)}, \hat{\mathbf{y}}^{(t)}) = -\sum_{w\in V}\mathbf{y}_w^{(t)} \log \hat{\mathbf{y}}_w^{(t)} = -\log \hat{\mathbf{y}}^{(t)}_{\mathbf{x}_{t+1}}.
    \end{aligned}
\end{equation}
By averaging the previous on the entire training set, we obtain the following overall loss:
\begin{equation}
    J(\theta)= \frac{1}{T}\sum_{t=1}^T  J^t(\theta)  = \frac{1}{T} \sum_{t=1}^T -\log \hat{\mathbf{y}}^{(t)}_{\mathbf{x}_{t=1}}.
\end{equation}

\textit{Embedding} representations play a crucial role in LMs. An embedding is a numerical representation of words, phrases, sentences, or even entire documents. These representations are typically high-dimensional vectors that capture the semantic meaning of the text. The importance of embeddings lies in their ability to transform text into a format that machine learning models can understand and process. Therefore, embeddings capture the nuanced meanings of words based on their context, which is essential for tasks like sentiment analysis, translation, and summarization. 

%
%

\section{Threat model}

\paragraph{Attack Description} A directory enumeration brute-force attack is a method that checks for and attempts to access directories and files on a web server that are not referenced by the application but are still accessible. This type of attack is performed by generating a large number of requests associated with different URLs sent to the server. The attack is commonly based on a wordlist, a list of words used to construct the URLs starting from the base one the attacker selects. 
\par
The primary goal of a directory enumeration attack is to uncover hidden files, directories, backup files, or administrative interfaces that may contain sensitive information or configuration data. If these resources are not adequately secured, they can be exploited to gain unauthorized access, escalate privileges, or launch further attacks. Vulnerabilities typically exploited by such attacks include misconfigured permissions, default installations with sample files, and outdated or unnecessary files left accessible on the server.
\par
Directory enumeration brute-force attacks are often employed during the \textit{reconnaissance phase} of a penetration test. A penetration test, or pentest, is an authorized simulated cyberattack on a computer system performed to evaluate its security. 
However, this type of attack may also be performed by malicious actors, so it is essential to be aware of this type of attack and to test web application security against it properly.

\paragraph{Automated tools}
Several commercial and open-source tools are commonly used to perform directory brute-force attacks. These tools can be specific to this type of attack or be more broad-based to provide other functionalities; additionally, they also often come with default wordlists. Popular tools are:
\begin{itemize}
\item \textit{Dirbuster}, a Java-based, multi-threaded tool specifically designed to brute force directories and file names on web or application servers developed by OWASP\footnote{\url{https://owasp.org/}}. It has nine different default wordlists. The tool is freely available at: \url{www.kali.org/tools/dirbuster/}.
\item \textit{Wfuzz}, an open-source security tool designed to launch brute-force attacks against web applications by fuzzing input parameters and assisting penetration testers in identifying vulnerabilities. It is designed to perform several attacks, such as brute-forcing, fuzzing, and injection attacks. It also comes with several wordlists covering a variety of contexts. The tool is freely available at: \url{wfuzz.readthedocs.io}.
\item \textit{Burpsuite}, a commercial platform that provides a graphical tool for conducting security testing on online applications. It supports the entire testing process, from initial mapping and analysis of an application’s attack surface to the discovery and exploitation of security flaws. Among these, Burpsuite can perform brute-force attacks to enumerate directories, given a target and a wordlist. The tool is available under different licences at: \url{portswigger.net/burp}.
\end{itemize}

\paragraph{Wordlists}
Wordlists are essentially a set of directories utilized in a brute-force attack. Therefore, they play a vital choice when using these tools. A proper choice of wordlist can greatly impact the results, potentially uncovering more vulnerabilities. 

In the context of directory brute-forcing attacks, there is a range of wordlist categories that fit different needs: from general-purpose wordlists to backup-file wordlists, CMS-specific (Content management system) wordlists, and even more.

Various automated tools are provided by default with various wordlists. However, many other user-created wordlists can be found on the Internet, and users may also create ad-hoc wordlists that satisfy their needs.
In the scope of this research, we selected four general-purpose wordlists to assess:
\begin{itemize}
\item \texttt{big\_wfuzz} [\textbf{BW}]\footnote{\url{https://github.com/xmendez/wfuzz/blob/master/wordlist/general/big.txt}}: a Wfuzz default general-purpose wordlist that contains 3024 words. 
\item \texttt{top\_10k\_github} [\textbf{GH}]\footnote{\url{https://github.com/xajkep/wordlists/blob/master/discovery/top-10k-web-directories_from_10M_urlteam_links.txt}}: a user-created wordlist in GitHub containing 10000 words, created selecting the most common words found in ten million URLs. 
\item \texttt{megabeast\_wfuzz} [\textbf{MW}]\footnote{\url{https://github.com/xmendez/wfuzz/blob/master/wordlist/general/megabeast.txt}}: another Wfuzz default general-purpose wordlist that contains 45459 words.
\item \texttt{directory-list\_dirbuster} [\textbf{DB}]\footnote{\url{https://github.com/3ndG4me/KaliLists/blob/master/dirbuster/directory-list-1.0.txt}}: a Dirbuster default wordlist containing 141835 words. 
\end{itemize}
\section{Methodology}

\paragraph{Overview}
Traditional attacks are essentially inefficient, as they are based on brute-forcing mechanisms.  
In this work, we explore two different approaches that might improve the attack: one based on probabilities and one using a Language Model for path generation.
Given the scope of this research, which aims to be general and not to focus on specific technologies or sensitive information, both approaches aim to highlight the feasibility of implementing more efficient attacks and aim to exploit two features not used by the traditional wordlist-based brute-forcing approach:
\begin{itemize}
\item \textit{\textbf{Prior Knowledge}}. Web applications that belong to similar categories might have a similar structure. Given a target website, using knowledge retrieved from similar websites to decide what HTTP requests to send to the target website may positively impact the results.
\item \textit{\textbf{Adaptive decision-making}}. During a directory brute-force attack, having the ability to dynamically decide which URLs to generate and which requests to send might improve the hit rate of successful responses and reduce ineffective requests. 
\end{itemize}

\paragraph{Tree reconstruction}
Before we discuss how traditional tools and our proposed approaches work, it is helpful to understand how HTTP requests allow us to reconstruct the filesystem of a web application. 
Since a filesystem has a hierarchical tree structure, the paths of each web application can be used to reconstruct it. In particular, we used the \texttt{AnyTree} class in Python to reconstruct the filesystems of each web application, considering as root the starting URL usually referred to as the target. This strategy allows us to perform depth-level analysis and simulations of offline brute-force attacks so that we do not perform actual attacks on online web applications, thus maintaining an ethical posture that still allows us to obtain meaningful results.

\begin{figure}[!htpb]
    \centering
    \includegraphics[width=0.8\columnwidth]{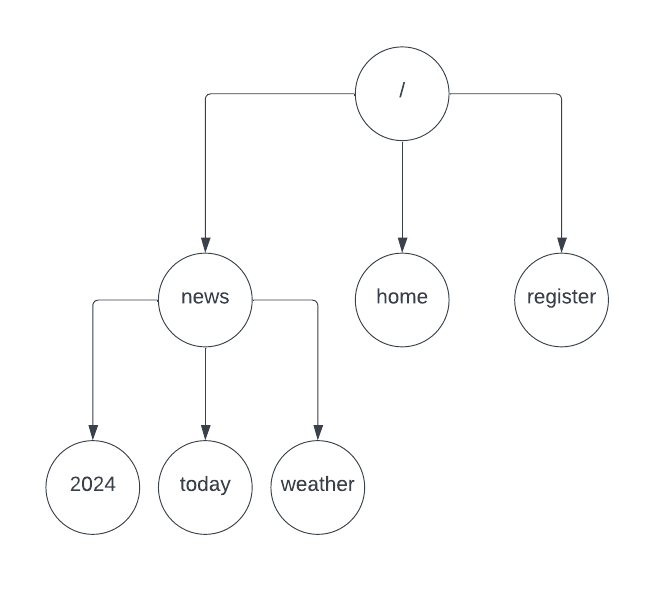}
    \caption{Visualization of a reconstructed tree.}
    \label{fig:reconstructedTree}
\end{figure}

For example, considering the paths "\texttt{/news}", "\texttt{/home}", "\texttt{/register}", "\texttt{/news/2024}", "\texttt{/news/today}" and "\texttt{/news/weather}" as the paths extracted from the crawl of a web application, we can visualize the corresponding reconstructed tree in Figure~\ref{fig:reconstructedTree}.

\subsection{Standard approach}\label{ssec.standard}
The standard wordlist-based approach that we will use to compare the results with our proposed approaches is based on two main strategies: \texttt{Depth-First} and \texttt{Breadth-First}.
\paragraph{Depth-First}
In a directory brute-force attack, the Depth-First approach prioritizes the exploration of subdirectories within a discovered directory before moving on to other directories at the same level. The algorithm initiates by sequentially sending HTTP requests using the entries in a wordlist. Upon receiving a positive response, which indicates the construction of a valid URL and, hence, the discovery of a valid directory, the algorithm shifts its focus to brute-forcing the subdirectories of this newly discovered directory. It exhaustively searches within these subdirectories before it resumes brute-forcing other directories at the same depth as the previously validated one. This approach ensures a comprehensive search within each directory before moving on to the next, thereby maximizing the chances of uncovering valuable information nested deep within the directory structure.
\par
An algorithmic representation of this approach can be visualized in Algorithm~\ref{alg:depth-first}, where \texttt{constructURL(URL, word)} is a function to generate a valid URL appending a word to the path of the URL and \texttt{isValid(response)} is a function that checks if the response is valid. 
\begin{algorithm}
    \caption{Depth-First brute-force attack Pseudocode}
    \label{alg:depth-first}
    \begin{algorithmic}[1]
    \Procedure{DepthFirst}{$rootURL, wordlist$}
    \For{each $word$ in $wordlist$}
    \State $url \gets$ constructURL($rootURL, word$)
    \State $response \gets$ sendHTTPrequest($url$)
    \If{isValid($response$)}
    \State DepthFirst($url, wordlist$) 
    \EndIf
    \EndFor
    \EndProcedure
    \end{algorithmic}
\end{algorithm}

\paragraph{Breadth-First}
In contrast to the previous approach, the Breadth-First approach prioritizes the exploration of directories at the same level before delving into their subdirectories. The algorithm begins by sending HTTP requests sequentially using the wordlist entries. When it receives a positive response, indicating the formation of a valid URL and, hence, the discovery of a valid directory, it continues to brute-force the remaining directories at the same depth. Only after it has exhausted all directories at the current level does it proceed to brute-force the subdirectories of the discovered directories. This method ensures a thorough search across each level of directories before descending deeper into the directory structure, thereby maximizing the chances of uncovering valuable information distributed across the directories. The majority of commercial tools implement this approach.
We also present a pseudocode implementation in Algorithm~\ref{alg:breadth-first}, using a \texttt{queue} to store and retrieve the URLs used during the process.
\begin{algorithm}
\caption{Breadth-First brute-force attack Pseudocode}
\label{alg:breadth-first}
\begin{algorithmic}[1]
\Procedure{BreadthFirst}{$rootURL, wordlist$}
\State $queue \gets$ new Queue()
\State $queue$.enqueue($rootURL$)
\While{$queue$ is not empty}
\State $currentURL \gets queue$.dequeue()
\For{each $word$ in $wordlist$}
\State $url \gets$ constructURL($currentURL$, $word$)
\State $response \gets$ sendHTTPrequest($url$)
 \If{isValid($response$)}
\State $queue$.enqueue($url$)
\EndIf
\EndFor
\EndWhile
\EndProcedure
\end{algorithmic}
\end{algorithm}

\subsection{Probability-based approach}\label{ssec.prob}
\paragraph{Approach}
Prior knowledge might be essential to improve the attack performance. The intuition is straightforward: if the majority of websites contain paths like \texttt{login} and \texttt{register}, it is likely that the tested website contains such directories as well. 
An algorithm that, therefore, prioritizes directories based on prior knowledge can be effective. 
\par
The first strategy we present optimizes the depth and breadth-first strategies described in Section~\ref{ssec.standard}, where the wordlist is ordered according to prior knowledge (e.g., gathered from web applications similar to the victim).
This adds dynamic decisions on how to go about generating the following HTTP request to maximize the number of positive requests while minimizing the number of unlikely and incorrect requests. 
The prior knowledge, or training dataset, contains crawls of paths of various web applications, possibly of the same category as the target where the attack will be performed.
The prior knowledge can be then infused into the algorithms in two possible manners:
\begin{enumerate}
\item Constructing a \textbf{Weighted Training Tree}. Using the same reasoning with which we described how it is possible to reconstruct a filesystem of a web application from the crawl of its paths, we will proceed to construct a single filesystem tree that unites all the paths that contain our training dataset indistinctly from the web application. Furthermore, this tree will be weighted: for each new node in the tree (corresponding to a directory), we would maintain a counter indicating how many times that particular node is repeated. An example of this is reported in Figure~\ref{fig:weightTree}.

\begin{figure}[!htpb]
    \centering
    \includegraphics[width=0.8\columnwidth]{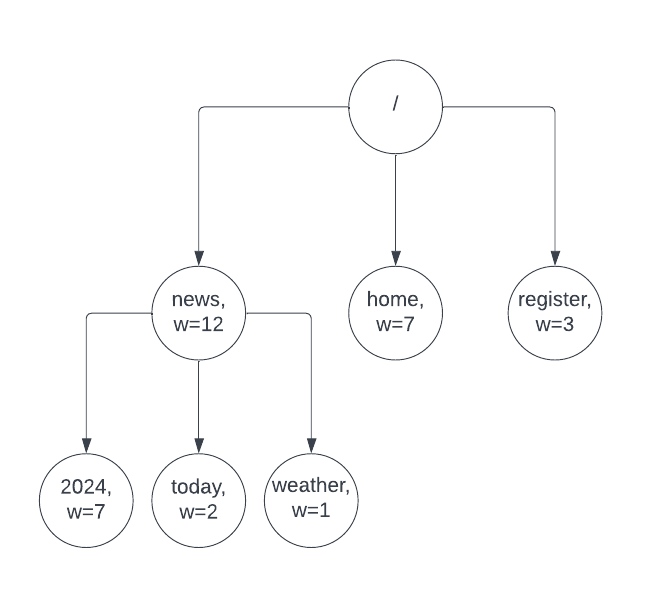}
    \caption{Visualization of a Weighted Training Tree, obtained merging paths from a Training Dataset.}
    \label{fig:weightTree}
\end{figure}

\item Constructing a \textbf{Weighted Wordlist Tree}. A weighted tree using only the words from the wordlist. Starting from a general wordlist, we construct a weighted tree similar to the Weighted Training Tree. However, this tree only includes words from a pre-designed wordlist (e.g., big\_wfuzz). The weight of each node (directory) in this tree is determined based on the training set. For example, if we consider the wordlist ["\textit{news}", "\textit{home}", "\textit{2024}", "\textit{today}", "\textit{about}"] and the Weighted training tree shown in Figure~\ref{fig:weightTree}, the corresponding Wordlist Weighted Tree can be visualized in Figure~\ref{fig:wordlistTree}. Note that, in this case, folders such as \texttt{register} and \texttt{weather} are not included in the tree, since they are not contained in the original wordlist.  

\begin{figure}[!htpb]
    \centering
    \includegraphics[width=0.8\columnwidth]{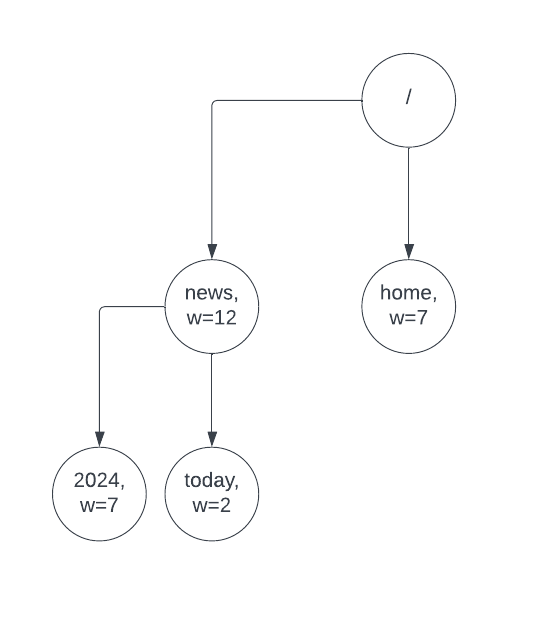}
    
    \caption{Visualization of a Wordlist Weighted Tree, based on a wordlist and a Weighted training tree.}
    \label{fig:wordlistTree}
    
\end{figure}
\end{enumerate}
In this way, we can make the best use of parent-child relational information between directories and subdirectories and give weight to words in the wordlist critical for the adaptive selection of requests to be made. 
In addition, an interested pruning process takes place of all those words in the wordlists that, for each directory, are not found to be valid subdirectories and, therefore, less likely than others. The pruning of unlikely words helps, consequently, to minimize the number of less probable requests. To give an example of how pruning works, considering Figure~\ref{fig:weightTree} and Figure~\ref{fig:wordlistTree}, we can see that the word "\texttt{about}", which was initially in the wordlist from which the tree is created, is not part of the Wordlist weighted tree (so it has a weight of 0 as a possible subdirectory for every directory), and "\texttt{news}" has a weight of 12 as a subdirectory of the base root, but is pruned from being a child-node of "\texttt{home}".


\paragraph{Algorithm}
The probabilistic approach employs a max heap (a data structure that keeps the maximum element of a given property on top of it) with tuples of base URLs, a word, and the weight assigned to it. The max heap keeps the tuples ordered by probability, so we can always pop the highest one to construct and send a request. 

The probability of a word being a valid subdirectory of a directory is computed dynamically by dividing the weight of the possible subdirectory by the sum of the weights of every possible subdirectory to that directory.

In the beginning, given a target base URL, the algorithm will push in the max heap all the possible subdirectories of the root directory "/" retrieved from the weighted tree with the corresponding probabilities. 
Whenever we receive a successful response, the algorithm pushes all the possible subdirectories to the response URL with the probabilities into the heap.
This mechanism allows us to implement an adaptive decision-making strategy to consistently send the most likely HTTP request away as new subdirectories are discovered.

An algorithmic exemplification of this approach is presented in the algorithm~\ref{alg:Probability}, where \texttt{getWordswithWeights()} returns the word-weight pairs taken from the specified URL in the weighted tree, and \texttt{getProbability()} calculates the probability of a word based on as described previously.
\begin{algorithm}[h]
\caption{Probabilistic brute-force attack Pseudocode}
\label{alg:Probability}
\begin{algorithmic}[1]
\Procedure{Probabilistic}{$rootURL, weightedTree$}
\State $maxHeap \gets$ new MaxHeap()
\State $rootTuples \gets$ getWordswithWeights ($rootURL$, $weightedTree$)
\For{each $word, weight$ in $rootTuples$}
\State $prob \gets$ getProbability($word$, $weight$ $rootTuples$)
\State $maxHeap$.push(($rootURL$, $word$, $probability$))
\EndFor
\While{$maxHeap$ is not empty}
\State $currentURL, word, probability \gets maxHeap$.pop()
\State $url \gets$ constructURL($currentURL$, $word$)
\State $response \gets$ sendHTTPrequest($url$)
\If{isValidURL($response$)}
\State $newTuples \gets$ getWordswithWeights($url$, $weightedTree$)
\For{each $word, weight$ in $newTuples$}
\State $newProb \gets$ getProbability($word$, $weight$ $newTuples$)
\State $maxHeap$.push(($url$, $word$, $newProb$))
\EndFor
\EndIf
\EndWhile
\EndProcedure
\end{algorithmic}    
\end{algorithm}

Because of aggressive pruning on the wordlist tree or due to a training dataset that does not contain as much data, the possible requests provided by the weighted tree will be exhausted quickly, even before reaching any set request budget. 
In light of this, when all potential requests derived from the existing knowledge within the weighted tree have been explored, it is advisable to employ a conventional breadth-first strategy. This approach takes into account the previously successful responses, thereby mitigating the need to reissue redundant HTTP requests.

\subsection{Language-Model based approach} \label{method:LM}
\paragraph{Approach}
Following the intuition of the probabilistic approach described in Section~\ref{ssec.prob}, we design a neural network mechanism leveraging Language Models (see Section~\ref{sec.background}) to generate probable subdirectories to a given path.
Given a URL, we can consider its path a sequence of words separated by "/." This sequence of words can be fed as input to the neural network mechanism, which will output the words that most likely follow the input sequence. Those words can be used to construct new URLs and send new HTTP requests.
\par
With this method, we aim to leverage the power of customized embeddings (i.e., embeddings trained on the corpus), and overcome the limitations of the probabilistic approach. 
In particular, the probabilistic approach calculates relationships among directories that appear in the prior knowledge. On the other hand, with the embedding, the model learns the context, and therefore directories appearing in a similar context will be used to generalize the attack. For instance, suppose that in our prior knowledge, we have URLs such as \texttt{"/account/setting/info"}, \texttt{"/account/setting/password"}, \texttt{"/account/setting/logout"},  \texttt{"/profile/setting/ password"}  
 and \texttt{"/profile/setting/info"}:
the directories \texttt{account} and \\ \texttt{profile} are utilized in a similar context, and therefore their embedding will be close. At inference time, a LM might infer the URL \texttt{"/profile/ setting/password"} even though this information was not available in our prior knowledge. In this example, a probability approach would have assigned 0 to this association.  

\paragraph{Model architecture}
Leveraging language models in our architecture involves using a crucial component of them: the vocabulary. This component maps words in our sequences (i.e., directories) to unique indexes.
The vocabulary allows the translation of words into integer indices that the neural network architecture can process.
To reduce the dimension of the vocabulary, only words more frequent than a certain threshold are considered.
 Additionally, vocabulary helps the architecture understand the structure of the sentences while handling unknown words and variable-size sequences with special tokens. These tokens are: \texttt{UNK} (Unknown word), \texttt{PAD} (Padding token, used to pad sequences to a fixed size), \texttt{SOS} (Start of sentence token, used to highlight where a sequence start) and \texttt{EOS} (End of sentence token, used to highlight where a sequence end).

Our designed neural network architecture is primarily based on the Long-Short-Term Memory (LSTM) network~\cite{hochreiter1997long}, a type of Recurrent Neural Network.
Our proposed LM architecture consists of several key components:
\begin{enumerate}
\item \textbf{Embedding Layer}. The first layer of the model is an embedding layer, which transforms the input words into dense vectors of fixed size, defined as embedded size. Embedding representations are learned at training time.
\item \textbf{LSTM Layer}. It follows a LSTM layer, crucial for the learning of patterns in sequential data.
This layer takes the embeddings of the input words and returns its own hidden states and cell states.
\item \textbf{Dropout Layer}. To prevent overfitting, dropout layers are used after the Embedding and LSTM layers. \textit{Dropout} is a regularization technique that randomly sets a fraction of input units to 0 with a specific frequency of rate at each step during the training.
\item \textbf{Fully Connected Layer}: The LSTM outputs (hidden states) are then passed through a fully connected (linear) layer to transform them into the desired output shape, which is the size of the vocabulary. 
\item \textbf{Softmax Function}: A softmax function is applied to transform the output of the fully connected layer to probabilities assigned to each vocabulary word.
\end{enumerate}

\begin{figure*}[!htpb]
    \centering
    \includegraphics[width=\textwidth]{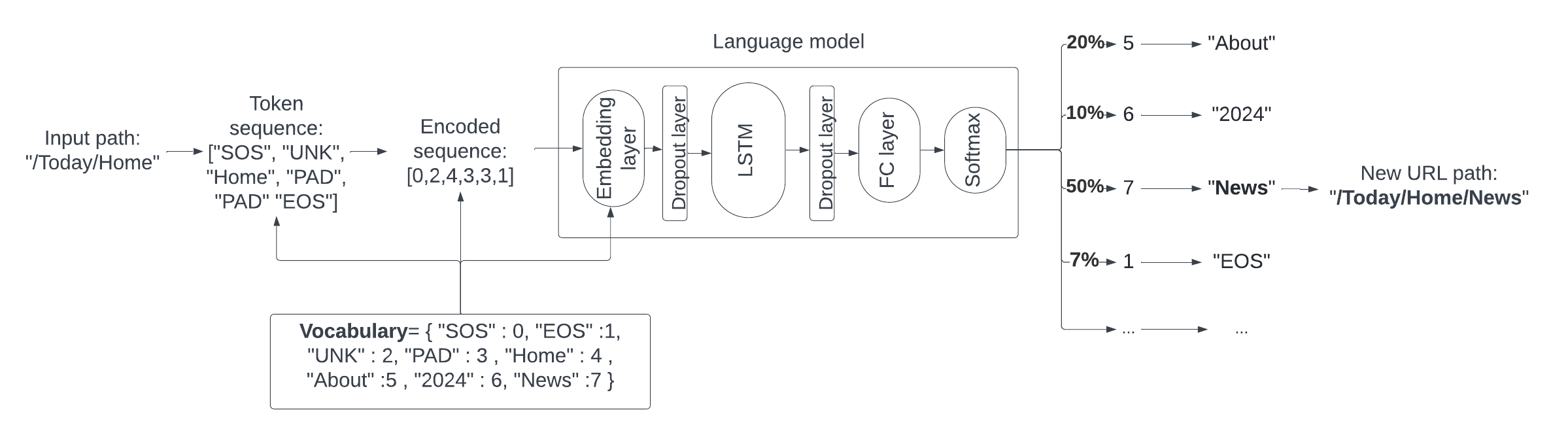}
    
    \caption{Prediction of the next directory from our LM-based architecture.}
    \label{fig:LM} 
\end{figure*}

Figure~\ref{fig:LM} shows an overview of the architecture. Here, we can see how the URL path is split into tokens and mapped into integers using vocabulary. Then, after the sequence of integers is given as input to the model, we observe how the softmax function takes the output of the fully connected layer and assigns the probability that it is the next in the sequence to each number. At this point, the most likely word is chosen to form a new path, but the choice can also be made on any other word.

\paragraph{Training and Validation} 
Our architecture's training process involves feeding it with paths and having it predict the following directory in the path. The model's predictions are compared to the actual following directories in the path, and a loss function is used to quantify the difference between the predictions and the truth. This loss is minimized using an optimization algorithm, which adjusts the model's parameters to make its predictions more accurate.


During the training phase, the model performance is periodically evaluated on the validation set to prevent overfitting on training data. This strategy allows us to monitor the model's generalization ability to unseen data. We utilize an \textit{early stopping} mechanism to stop the training when the model's performance on the validation set starts to deteriorate (a phenomenon known as overfitting) or does not improve for $p$ epochs.


\paragraph{Algorithm}
The algorithm incorporating the proposed neural network architecture (Algorithm \ref{alg:LM}) uses a strategy similar to Algorithm 3, where a max heap is used to construct the most probable URL. Instead of using the wordlist weighted tree, the language model in the function \texttt{predict()} is used to return a number of word-probability pairs with a higher probability specified in the \texttt{topPredicts} hyper-parameter.

\begin{algorithm}[h]
\caption{Language-model based brute-force attack Pseudocode}
\label{alg:LM}
\begin{algorithmic}[1]
\Procedure{LMattack}{$rootURL, LM, topPredicts$}
\State $maxHeap \gets$ new MaxHeap()
\State $rootTuples \gets$ predict($rootURL$, $LM$,$topPredicts$)
\For{each $word, prob$ in $rootTuples$}
\State $maxHeap$.push(($rootURL$, $word$, $prob$))
\EndFor
\While{$maxHeap$ is not empty}
\State $currentURL, word, prob \gets maxHeap$.pop()
\State $url \gets$ constructURL($currentURL$, $word$)
\State $response \gets$ sendHTTPrequest($url$)
\If{isValidURL($response$)}
\State $newTuples \gets$ predict($url$, $LM$,$topPredicts$)
\For{each $word, prob$ in $newTuples$}
\State $maxHeap$.push(($url$, $word$, $newProb$))
\EndFor
\EndIf
\EndWhile
\EndProcedure
\end{algorithmic}    
\end{algorithm}

\section{Dataset}
In this section, we present the datasets collected for our experiments. In particular, Section~\ref{ssec.dataset} describes the data collection process. It follows Section~\ref{ssec.data-analysis}, presenting an in-depth analysis of our data. 

\subsection{Description}\label{ssec.dataset}
\paragraph{Source of data}
The data for this research is obtained from CommonCrawl~\footnote{\url{https://commoncrawl.org/}}, a non-profit organization that crawls the web and freely provides its archives and datasets. CommonCrawl was selected due to its comprehensive repository of web crawls that are updated regularly; we utilized \texttt{CC-MAIN-2023-40} crawl version. 
This approach not only streamlines the data collection process but also aligns with ethical considerations, avoiding manual spidering and crawling of websites that may be categorized as brute-force attacks and not overloading web servers with severe amounts of requests but instead using historical data.
Given the scope of the search and the multitude of data in the CommonCrawl corpus, we collected the URLs from the HTTP responses. We then extracted the \texttt{domain}, \texttt{path}, and \texttt{response code} necessary to identify invalid responses.

\paragraph{Datasets}
Considering the increasing number of cyber attacks and the wide variety of possible targets, we decided to consider four different datasets representing some most common categories of organizations at risk of cyber attacks. In addition, given the scope of the search, we maintained only websites written in English. We now list the four distinct datasets we collected:  
\begin{itemize}
\item \textit{Universities dataset} [\textbf{UNI}]: consisting of the HTTP responses from 100 English-based web applications of the top universities listed in the QS 2023 World University Rankings~\footnote{\url{https://www.topuniversities.com/world-university-rankings/2023}}. Furthermore, we did not consider universities without an English version of their web application.
\item \textit{Hospitals dataset} [\textbf{HOS}]:  consisting of the HTTP responses belonging to the web applications of the first USA 100 hospitals listed in "Ranking Web of World Hospitals"~\footnote{\url{https://hospitals.webometrics.info/en/americas/usa}}. 
\item \textit{Companies dataset} [\textbf{COM}]: consisting of the HTTP responses from 100 corporate web applications of companies in the S\&P 500~\footnote{\url{https://www.slickcharts.com/sp500}}, choosing in order of highest capitalization (January 2024) and avoiding companies that had e-commerce as their main web application.
\item \textit{Government dataset} [\textbf{GOV}]: consisting of the HTTP responses from 336 different USA government web applications~\footnote{\url{https://www.usa.gov/agency-index}}.
\end{itemize}
In our experiments, we will report the attack performance when considering the datasets separately and together. 

\paragraph{Preprocessing}
After extracting the domain, path, and status code from each HTTP response for each dataset, we performed additional preprocessing steps on the data to reconstruct the hierarchical structure of the file system of the Web applications, guaranteeing their integrity and enabling more consistent analysis in the datasets.
Firstly, we applied initial filtering to the HTTP responses, preserving only the responses with a status code 200, representing most of the crawled responses. The HTTP status code 200 indicates that the client request has succeeded.
Secondly, eventual queries or files that were part of the path were removed. This choice was mainly made to maintain a manageable scope of analysis. URLs can often contain queries or files with different extensions that introduce a significant degree of variability and are highly dependent on the technology with which the web application was developed, thus going against our intended general approach.
Additionally, it is useful to define the \textit{depth} of a path: considering a path as a sequence of words (where each word corresponds to the name of a directory), we define the \textit{depth} as the number of words that the path consists of. The order matters since each directory in that path will be at a given depth (e.g.: "\texttt{/news/2023}" will have depth = 2, where "\texttt{news}" is at depth 1, and "\texttt{2023}" is at depth 2).

\subsection{Datasets Analyses}\label{ssec.data-analysis}

\paragraph{Overview}
Following the data preprocessing, we analyzed the characteristics of each dataset and the similarities and differences between them, which allowed us to get a general overview of the datasets and highlight what direction the test results might take.

We conduct the following analyses:
\begin{itemize}
    \item \textit{Dataset description}, where we describe datasets' properties (e.g., quantities, distributions) and the nature of their URLs. 
    \item \textit{Wordlist Coverage Ratio Analysis}, where we describe how standard wordlists can cover the retrieved URLs. 
    \item \textit{Stemming Analysis}, where we attempt to understand the impact of small name variations in the directories (e.g., \texttt{books} and \texttt{book}) and wordlists coverage.  
    \item \textit{Dataset Similarity Analysis}, where we describe the degree of similarity between directories in the four collected datasets (e.g., how universities and hospital URLs differ).     
\end{itemize}

\subsubsection{Dataset description}
We now describe the four distinct datasets we retrieved. 
For each dataset, we analyzed the following information: 
\begin{itemize}
    \item Number of domains (\texttt{\# Domains}). 
    \item Number of paths in each dataset (\texttt{\# Paths}). For instance, suppose a dataset contains two web apps, each with one domain (e.g., "\texttt{domain1/login}" and "\texttt{domain2/login}"), the number of paths is two, i.e., \texttt{["/login", "/login"]}).   \item The average number (and standard deviation) of paths of the web apps contained in a given dataset (\texttt{\# paths AVG} and {\# paths STD}). For instance, given two web apps containing each 1 URLs, the average is equal to 1, and the standard deviation to 0. 
    \item The number of unique paths in the dataset (\texttt{\# U-Paths}). For instance, suppose the dataset contains the following samples " \texttt{domain1/account/info} " and " \texttt{domain2/account/ info}", the unique paths are defined as the unique directory \texttt{["/account/info"]}.
    \item The number of directories (\texttt{\# Dir}) contained in the dataset. For instance, given "\texttt{domain1/account/info}" and "\texttt{domain2 /account/settings}", the dataset contains four directories ["\texttt{account}", "\texttt{info}", "\texttt{account}", "\texttt{settings}"].
    \item The number of unique directories (\texttt{\# U-Dir}) contained in the dataset.
    For instance, given "\texttt{domain1/account/info}" and "\texttt{domain2/account/settings}", the dataset contains three unique directories ["\texttt{account}", "\texttt{info}", "\texttt{settings}"].
    \item We analyze the depth of URLs in terms of average and standard deviation (\texttt{Depth AVG} and \texttt{Depth STD}). For instance, the URL "\texttt{domain1/account/info}" has a depth equal to two.
    Hypothesizing that a directory with greater depth is likely to be more specialized and, consequently, less common, analyzing depth distribution might provide insight into a web application's granularity and structure.
    \item The average depth and standard deviation of URLs similarities in a given dataset (\texttt{Sim AVG} and \textit{Sim STD}). In more detail, this metric computes the similarity for each pair of websites in each dataset. By defining a website as a set of its own directories (computed by the retrieved URLs), the similarity between two websites can be computed using the Jaccard Similarity, defined in Equation~\ref{eq.jaccard}. This metric is defined between 0 and 1, where 0 means that two sets are distinct, while one means identical.
\end{itemize}
\begin{equation}\label{eq.jaccard}
    J(A, B) = \frac{|A \cap B|}{|A \cup B|}.
\end{equation}
Table~\ref{tab:stats} shows the statistics for each dataset.


\begin{table}[!htpb]
    \centering
    \small
    \begin{tabular}{c|ccccc} \toprule 
        & \multicolumn{4}{c}{\textit{Dataset}} \\ \midrule
         \textit{features} & \textbf{UNI} & \textbf{HOS} & \textbf{COM} & \textbf{GOV}\\ \midrule 
         \# Domains & 88 & 80 & 97 & 336 \\
        \# Paths & 209657 & 211911 & 147198 & 520571 \\
        \# Paths AVG & 2301 & 2584 & 1479 &  1507\\
        \# Paths STD  & 2906 & 2800  &  2360 & 2340  \\
        \# U-Paths & 201768 & 205587 & 143067 & 502693 \\
        \# Dir & 203613 & 209945 & 143620 & 512595 \\
        \# U-Dir & 171215 & 173394 & 106097 & 462812 \\
        Depth AVG & 4.11 & 3.31 & 4.43 & 3.40 \\
        Depth STD & 1.69 & 1.72 & 2.23 & 1.66 \\
        Sim AVG & 0.022 & 0.019 & 0.016 & 0.016 \\
        Sim STD & 0.031 & 0.017 & 0.023 & 0.024 \\ \bottomrule
    \end{tabular}
    \caption{Summary statistics for the four datasets: universities [\textbf{UNI}], hospitals [\textbf{HOS}], companies [\textbf{COM}], and government [\textbf{GOV}].}
    \label{tab:stats}
\end{table}


\paragraph{Discussions}
It is interesting to observe that websites have different structures in terms of number of pages. For instance, universities and hospitals tend to have a bigger number of pages (\texttt{\# Path AVG}) compared to companies and governments. 
From a depth perspective, university and company websites tend to have deeper web app structures compared to hospitals and government. 
Last, the similarity analysis clearly shows that websites in the same dataset tend to have different types of directories, and the average Jaccard similarity tends to zero. 
By considering these statistics together, we can clearly remark why directory enumeration is not trivial, and it might require an enormous amount of requests for a small number of hits (i.e., discovered directories).

\subsubsection{Wordlists Coverage Ratio Analysis}
\begin{figure*}[!htpb]
\centering
\begin{subfigure}[h]{0.24\linewidth}
\includegraphics[width=\linewidth]{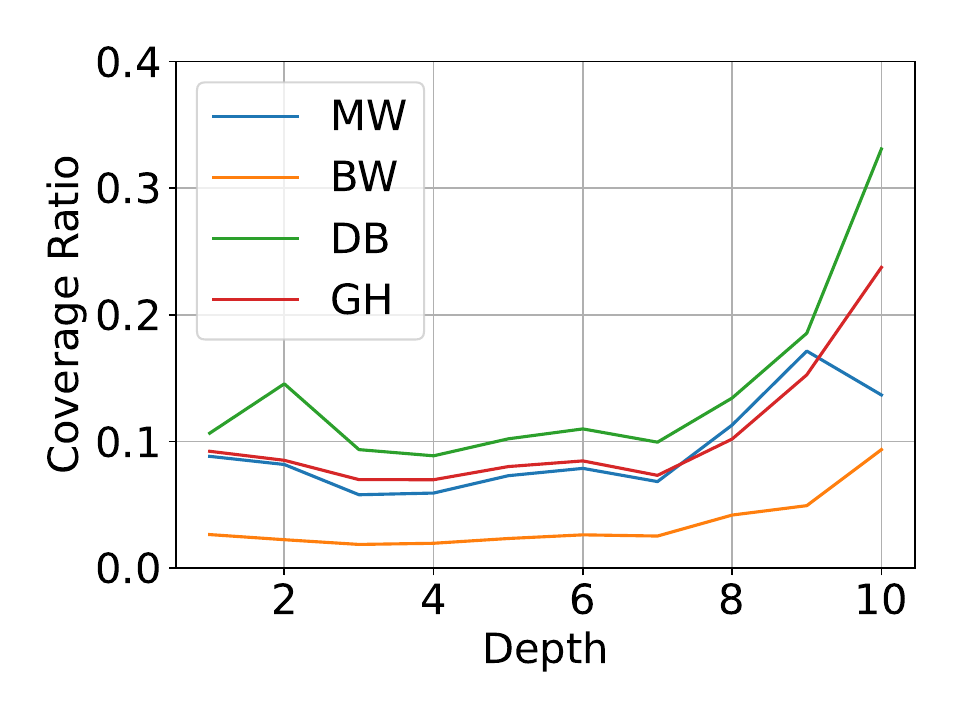}
\caption{University.}
\end{subfigure}
\hfill
\begin{subfigure}[h]{0.24\linewidth}
\includegraphics[width=\linewidth]{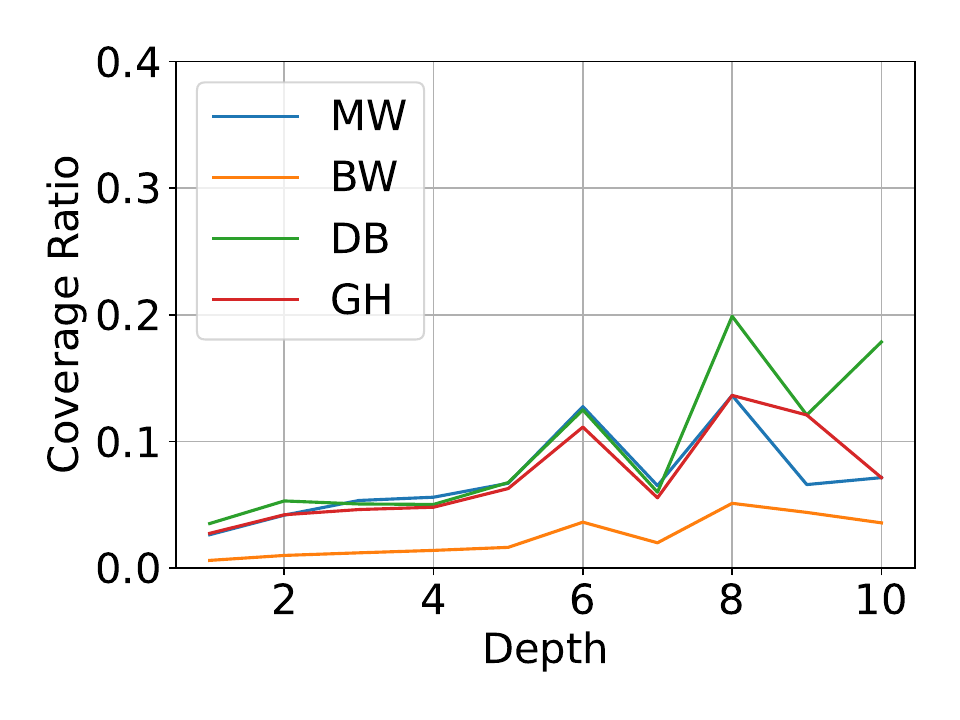}
\caption{Hospital.}
\end{subfigure}
\begin{subfigure}[h]{0.24\linewidth}
\includegraphics[width=\linewidth]{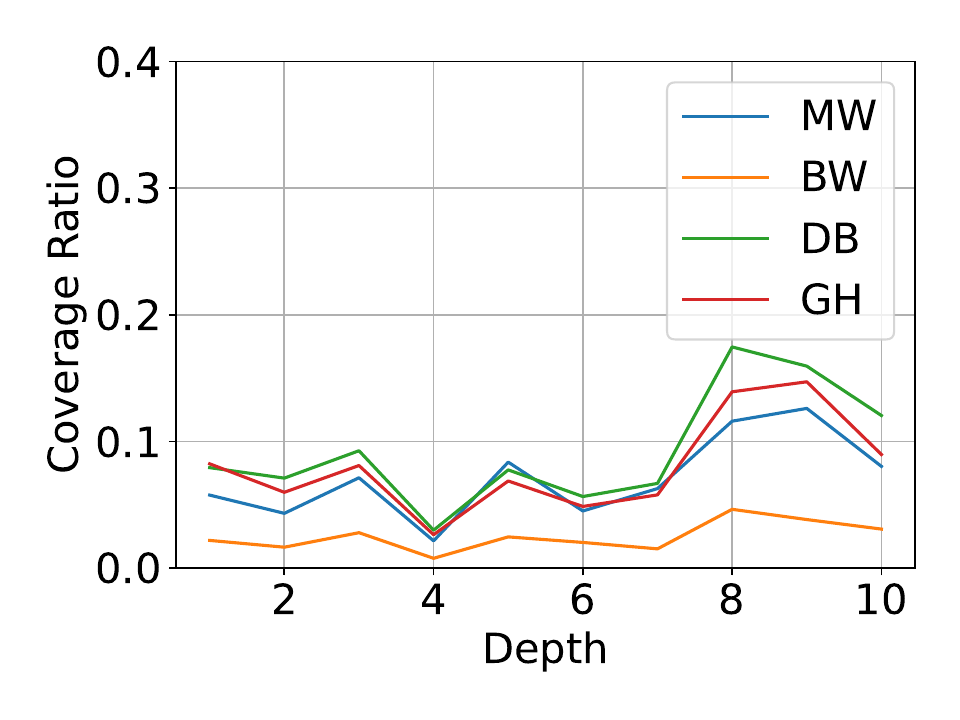}
\caption{Company.}
\end{subfigure}
\begin{subfigure}[h]{0.24\linewidth}
\includegraphics[width=\linewidth]{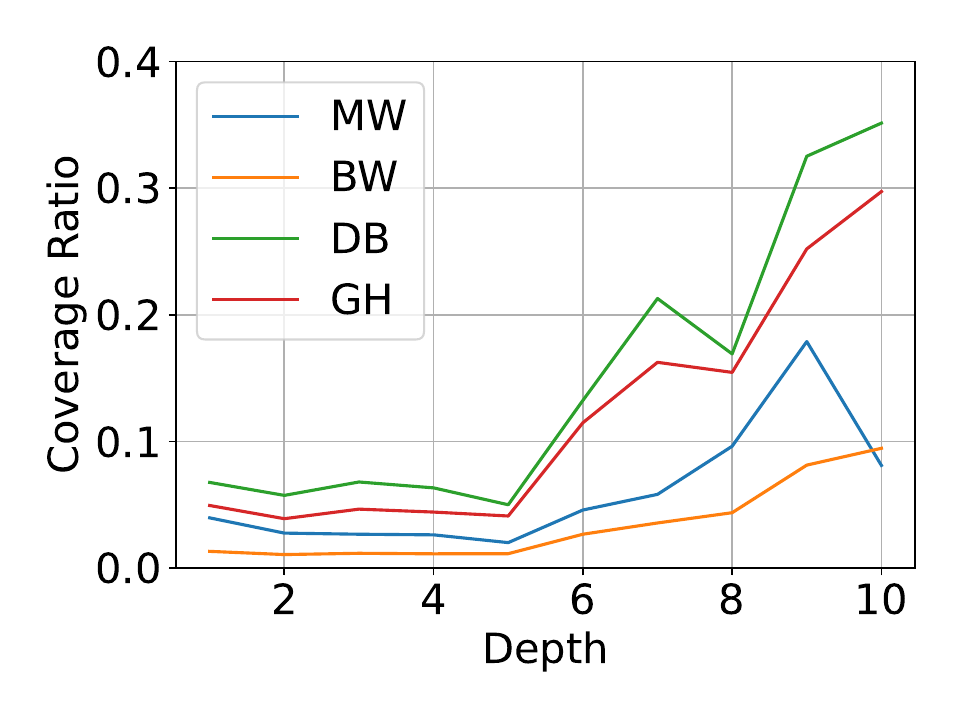}
\caption{Government.}
\end{subfigure}
\caption{Coverage analysis at the varying of the four datasets and four wordlists: ADD.}
\label{fig:coverage}
\end{figure*}

By analyzing the coverage of the different wordlists on the web applications in each dataset, shown in Figure~\ref{fig:coverage}, we can see that a low percentage of words are found even at low depths where we would expect them to be more common and thus present in the wordlist.

The \texttt{coverage ratio} across the different datasets shows considerable variance but overall low values, highlighting how directory brute-force attacks using these wordlists could potentially miss multiple valid requests.

Although the \texttt{coverage ratio} shows an upward trend as depth increases, it is essential to emphasize that although the higher-depth words might be more specific, their number is significantly reduced (as highlighted by the depth distribution analyzed earlier). In addition, the most critical point concerns the poor coverage of the initial words, which form the basis of most pathways: higher-depth directories will not be explored if antecedent ones are not explored.


\paragraph{Stemming Analysis}
Stemming is a linguistic process that simplifies words to their base or root form, known as the stem, often by removing common prefixes or suffixes. 
For example, stemming removes plural (dogs $\xrightarrow{}$ dog), -ing form (running $\xrightarrow{}$ run), etc. 
In our study, we employed the Porter-Stemmer~\cite{bird2009natural} algorithm to analyze the effect of stemming on the total number of unique words within our datasets.
In our analysis, we found a few instances of how a root form represents minimal variations of the same word, highlighting different conventions or singular and plural forms. A pair of examples are:
\begin{enumerate}
\item "\textit{articl}" corresponding to "\textit{article}" in 33.5\% of cases, "\textit{articles}" in 14.4\%, "\textit{Article}" in 47.3\%, "\textit{Articles}" in 4.7\%, and "\textit{ARTICLE}" in 0.01\%.
\item "\textit{project}" corresponding to "\textit{project}" in 46.78\% of cases , "\textit{projects}" in 53.13, "\textit{Projected}" in 0.04\% and "\textit{Projects}" in 0.04\%.
\end{enumerate}
However, these represent only a minority of cases, as most root forms correspond to only one word, and the percentage reduction in the datasets remains marginal, as shown in Table~\ref{tab:stemming}.
These statistics highlight how the words that make up our directory list differ (although some words have various declinations) and how that should be taken into account when designing new, improved approaches.


\begin{table}[h]
\centering
\small
\begin{tabular}{c|cccc}
\toprule
& \multicolumn{4}{c}{\textit{Dataset}} \\ \midrule
         \textit{features} & \textbf{UNI} & \textbf{HOS} & \textbf{COM} & \textbf{GOV}\\ \midrule
\begin{tabular}[c]{@{}l@{}} \# U-Dir \end{tabular} & 171215 & 173394 & 106097 & 462812 \\ 
\begin{tabular}[c]{@{}l@{}}\# U-Root\end{tabular} & 168462 & 171371 & 104220 & 457912 \\ 
Reduction & 2753 & 2023 & 1877 & 4900 \\
Reduction (\%) & 1.61\% & 1.17\% & 1.77\% & 1.06\% \\

\end{tabular}%
\caption{Summary statistics after the STEMMING for the four datasets: universities [\textbf{UNI}], hospitals [\textbf{HOS}], companies [\textbf{COM}], and government [\textbf{GOV}].}
\label{tab:stemming}
\end{table}

\begin{figure}[!htpb]
\begin{subfigure}[h]{0.7\linewidth}
\includegraphics[width=\linewidth]{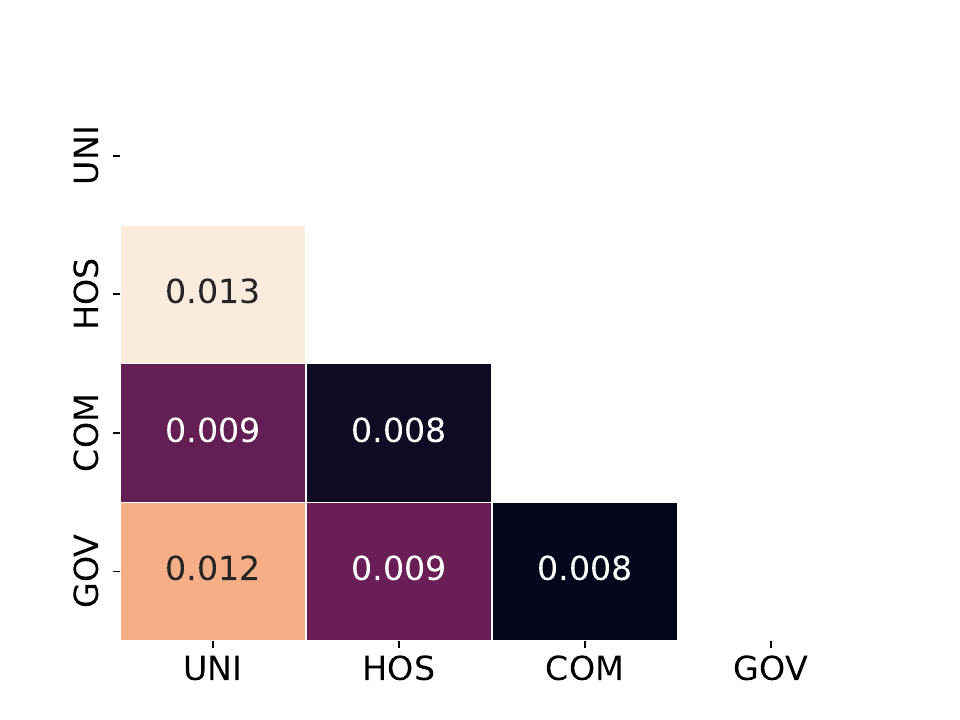}
\caption{Jaccard similarities between datasets wordlist.}
\end{subfigure}
\hfill
\begin{subfigure}[h]{0.7\linewidth}
\includegraphics[width=\linewidth]{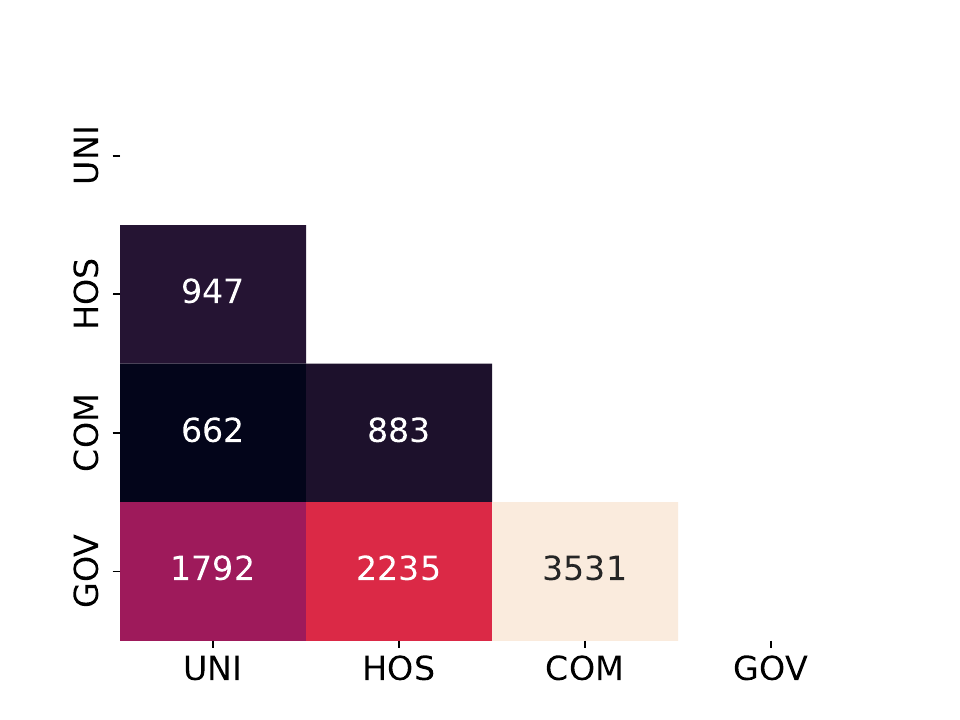}
\caption{Number of common paths.}
\end{subfigure}%
\caption{Similarity analysis for the four dataset: universities [\textbf{UNI}], hospitals [\textbf{HOS}], companies [\textbf{COM}], and government [\textbf{GOV}].}
\label{fig:similarity}
\end{figure}

\paragraph{Similarity Analysis}
Last, we measure the similarity between any pair of the collected dataset. 
We utilize two metrics: the Jaccard similarity of each dataset wordlist, and the number of paths in common (relative number) between the datasets. Figure~\ref{fig:similarity} shows the results. The first clear outcome is highlighted by the low Jaccard similarities: each dataset contains different directories. In other words, how websites of universities have almost completely different structures compared to hospital ones. This reasoning can be applied to any pair of datasets we utilized. 
A second interesting outcome is given by the relative count of common directories among different datasets. For instance, government and company websites contain many common paths. 
Considering the number of unique directories shown in Table~\ref{tab:stats}, it is clear that it is non trivial to design an effective directory enumeration brute-force attack.

\section{Results}
\subsection{Experimental Settings}
\paragraph{Testbed}
To set up our brute-force directory attack simulations, we partitioned the datasets into distinct sets for training, validation, and testing. 
We train our proposed approaches (i.e., probabilistic and LM-based) in a training set that merges the four presented in Section~\ref{ssec.dataset}. Merging the four data sources for the training is essential to have a sufficient number of samples to train a LM with. 
We, therefore, merge the four datasets and divide them into training, validation, and testing sets with a 70-10-20 split ratio, as mentioned in Section~\ref{method:LM}.
Note that the split is not random in terms of URLs, but from a domain perspective. In this way, all URLs of a given website will appear only in training, validation, or testing set. With this approach, we avoid any \textit{data snooping}~\cite{arp2022and}.
\par
Regarding the testing environment, we opted to simulate brute-force attacks offline, utilizing the virtual filesystems reconstructed from the test applications. This approach permits executing multiple attack simulations using different strategies without actualizing real-time brute-force attacks on live web applications. Furthermore, this approach allows us to work with a high number of simulated requests without introducing latency due to the HTTP requests.
Additionally, aligning with our objective to maximize successful responses while minimizing request volume, we established a maximum budget of 100,000 requests spendable by each simulated attack before termination.

\paragraph{LM validation}
We utilize PyTorch~\cite{pytorch} to design our architecture.
LM uses the training set to learn meaningful association, while the validation set is utilized to select the best LM hyperparameters. 
For the model selection, we use a grid-search validation over the following hyper-parameters.
\begin{itemize}
    \item \textit{Data representation}. We identify hyper-parameters that controls the training input: the maximum length of the paths utilized in the training phase and the minimum frequency that a directory must have not to be discarded and marked as "Unknown." For the former, defined as \texttt{max\_depth}, we chose values [5, 10], while for the latter, defined as \texttt{min\_freq}, we chose [3, 5].
    \item \textit{LM architecture}. \texttt{embedding\_size} \textbf{[ES]} = [128, 256, 512], \texttt{n\_layers} \textbf{[NL]} (number of layers in the LSTM) = [2, 3, 4], \texttt{dropout\_rate} \textbf{[DR]}=[0.2, 0.4, 0.6].
\end{itemize}
An early stopping mechanism is set with patience equal to 10 epochs.
The learning phase uses Adam~\cite{adam} as optimizer, and CrossEntropy as loss function.
Finally, the best model is chosen based on the lower loss at the validation set. 
\par
In addition, we tested an additional parameter, namely the number of predictions to be considered whenever a positive response is received during the attack and new predictions are made. The parameter, defined as \texttt{topPredicts} in Algorithm~\ref{alg:LM}, is tested with the values  [100, 250, 500, 750, 1000, 2000, 5000, 10000].





\paragraph{Evaluation Metric}
We utilize the following evaluation metrics:
\begin{itemize}
    \item \textit{Average Successful Response Rate}, consisting of averaging the total amount of successfully discovered directories for each tested website.   
    \item \textit{Bins efficiency}, consisting of averaging the total amount of successfully discovered directories for each tested website in a given range of requests. We evaluate the following bins: 0-100, 101-1000, 1001-10000, 10001-50000, and 50001-100000.
\end{itemize}

We did not consider execution times as they depend on various factors (such as the number of threads used in the attack and time between request and response variable for each application) and are not quantifiable with offline simulations.

\subsection{Results}
\paragraph{Overview}
Table~\ref{tab:results} shows the overall results obtained in our experiments, at the varying of the dataset, wordlists, and inference techniques. LM-based attack outperforms all the other baselines in all datasets, demonstrating the superiority of language models compared to naive techniques like brute-force or probabilistic approaches. 
Interestingly, LM's performances are not equally balanced on all datasets: for instance, LM struggles with university and company websites, while being robust with hospital and government ones. 
The probabilistic-based approach further shows a good improvement over brute-force attacks since, with a limited budget, the latter might not be able to fully assess all the possibilities. On the other hand, probabilistic approaches, by optimizing the priorities of the requests, are more efficient.

\begin{table}[!htpb]
\centering
\resizebox{\columnwidth}{!}{%
\begin{tabular}{cccccc}
\hline
\multicolumn{1}{c|}{\textit{Wordlist}} & \multicolumn{5}{c}{\textit{Dataset}} \\ \hline
\multicolumn{1}{c|}{} & \textbf{UNI} & \textbf{HOS} & \textbf{COM} & \textbf{GOV} & \textbf{ALL} \\ \hline
\multicolumn{6}{c}{\textbf{Breadth}} \\ \hline
\multicolumn{1}{c|}{big\_wfuzz} & 28.0 & 22.0 & 27.0 & 35.0 & 35.0 \\
\multicolumn{1}{c|}{directory-list\_dirbuster} & 8.0 & 10.3 & 9.6 & 11.8 & 10.5 \\
\multicolumn{1}{c|}{megabeast\_wfuzz} & 10.5 & 11.6 & 11.0 & 12.4 & 11.7 \\
\multicolumn{1}{c|}{top\_10k\_github} & 21.3 & 42.6 & 26.8 & 27.0 & 28.6 \\ \hline
\multicolumn{6}{c}{\textbf{Depth}} \\ \hline
\multicolumn{1}{c|}{big\_wfuzz} & 28.0 & 22.0 & 27.0 & 33.0 & 33.0 \\
\multicolumn{1}{c|}{directory-list\_dirbuster} & 0.5 & 0.5 & 0.7 & 0.4 & 0.5 \\
\multicolumn{1}{c|}{megabeast\_wfuzz} & 2.6 & 2.9 & 2.7 & 2.7 & 2.7 \\
\multicolumn{1}{c|}{top\_10k\_github} & 10.1 & 10.1 & 10.1 & 10.0 & 10.1 \\ \hline
\multicolumn{6}{c}{\textbf{Probability}} \\ \hline
\multicolumn{1}{c|}{big\_wfuzz} & 28.0 & 22.0 & 27.0 & 34.5 & 34.5 \\
\multicolumn{1}{c|}{directory-list\_dirbuster} & 14.0 & 13.1 & 11.1 & 17.3 & 25.4 \\
\multicolumn{1}{c|}{megabeast\_wfuzz} & 12.5 & 13.9 & 11.8 & 13.8 & 13.8 \\
\multicolumn{1}{c|}{top\_10k\_github} & 23.4 & 42.9 & 26.1 & 27.1 & 26.7 \\
\multicolumn{1}{c|}{train-set} & 31.9 & 60.4 & 27.6 & 30.8 & 42.5 \\ \hline
\multicolumn{5}{c}{\textbf{LM}} \\ \hline
\multicolumn{1}{c|}{train-set} & \textbf{90.0} & \textbf{175.0} & \textbf{89.0} & \textbf{128.0} & \textbf{175.0} \\ \hline
\end{tabular}%
}
\caption{Average successful responses for each approach achieved for different test-sets at the varying of the datasets. In bold the best results.}
\label{tab:results}
\end{table}

\begin{figure}[!htpb]
    \centering
    \includegraphics[width= \columnwidth]{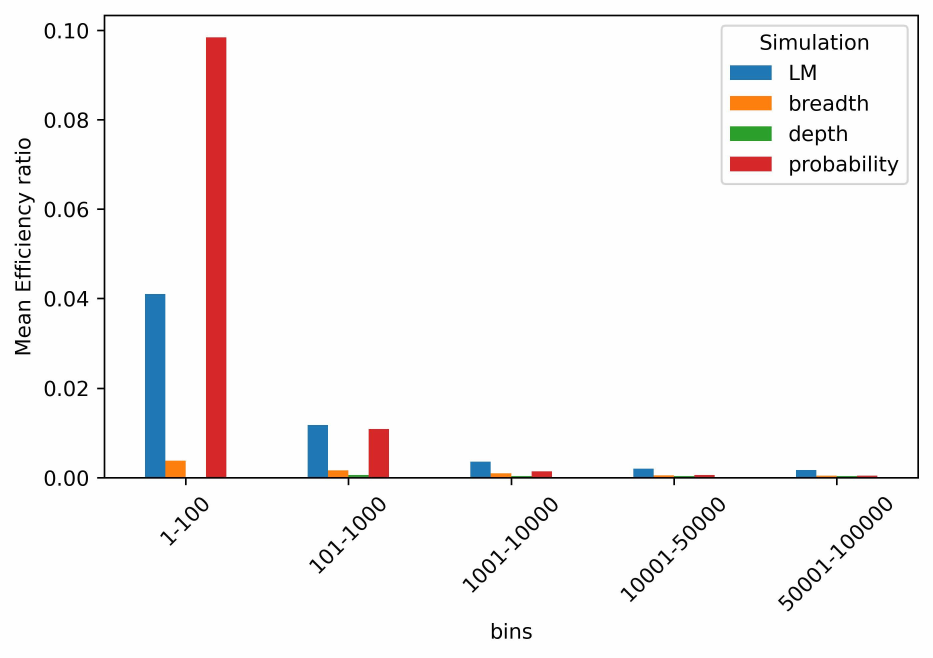}
    \caption{Mean Efficiency Ratio of the four approaches on different bins, considering wordlist = \texttt{big\_wfuzz} and the general dataset.}
    \label{fig:bins}
\end{figure}

\paragraph{Bins efficiency}
The efficiency analysis provides another interesting perspective on how different approaches perform. We analyze the efficiency shown in Figure~\ref{fig:bins}, calculated by considering simulations on the general test dataset (\textbf{[ALL]}) and on the wordlist \texttt{big\_wfuzz}. 
\par
Although the mean results obtained using the \texttt{breadth}, \texttt{depth}, and \texttt{probabilistic} approaches are the same, the efficiency varies considerably. The probabilistic approach performs very well in initial requests and then declines as requests increase.
Although the language model approach registers a 400\% increase in average successful responses, it performs worse in initial requests than the probabilistic approach but is more efficient in the long run.
This behaviour between the two is also observable in the other cases.
Therefore, adopting a probabilistic approach might be ideal when the budget is more limited. 

\subsection{Discussion}
\begin{figure}[!htpb]
    \centering
    \includegraphics[width= \columnwidth]{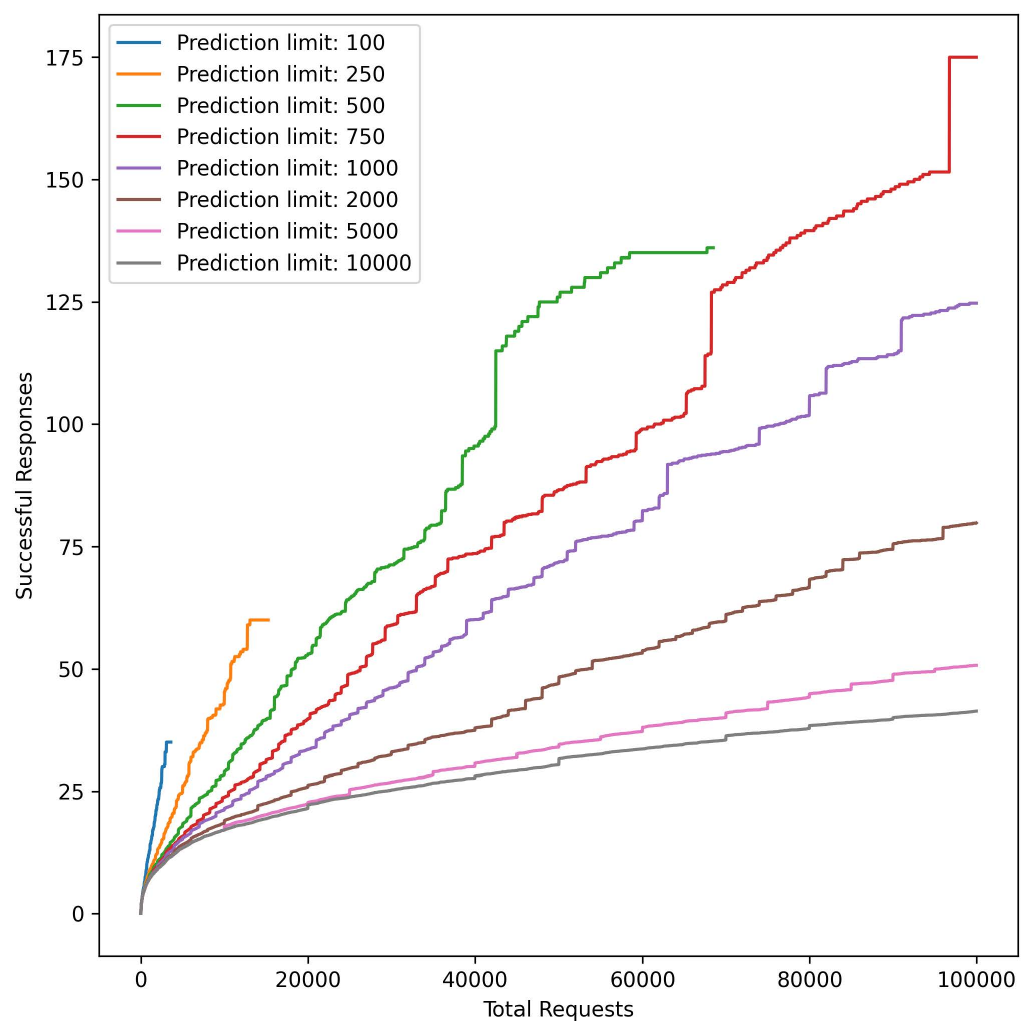}
    \caption{Evolution of average successful responses for different topPredicts values}
    \label{fig:Predictions}
\end{figure}

\paragraph{The impact of \texttt{topPredicts}}
Algorithm~\ref{alg:LM} relies on many hyper-parameters, such as \texttt{topPredicts}. 
It controls the number of most likely predictions to be considered in each new folder found. 
We analyze how the best LM found changes its performance at the varying of this setting. In particular, we explore the following: 100, 250, 500, 750, 1000, 2000, 5000, and 10000. 
As shown in Figure~\ref{fig:Predictions}, as the number of predictions considered increases, the average number of successful responses received in the attack simulations decreases. In addition, although with a smaller \texttt{topPredicts} the initial average number of successful responses received is better, the simulations end earlier as they have no more new predictions with which to send new requests. It is, therefore, essential to consider a value for this parameter that maximizes the results and utilizes the entire predetermined request budget.
Therefore, it might be ideal to set \texttt{topPredicts} to a small number if our budget is limited, as the model reaches the most successful responses in a short time. 
On the opposite, larger numbers like 500, 750, and 1000 are ideal when the budget allows an exhaustive search.

\paragraph{Results analysis}
The two proposed approaches show substantial improvements in both metrics examined.
Of the two standard approaches, the \texttt{breadth-first} strategy (also implemented by commercial tools) emerges as the best.

The probabilistic approach improves the performance of the breadth-first approach in 65\% of the cases considering the four default wordlists, while in the remaining, it obtains equal or slightly lower results. Depth-first approaches are outperformed in 100\% of the cases. 
In particular, if we consider the breadth-first approach, the probability-based approach using the train-set wordlist has the following average improvements: University +141\%, hospital +281\%, companies +85\%, government +78\%, and \textbf{ALL} +159\%.

The strength of this approach is the efficiency of successful responses received using few requests, which outperforms all other approaches considerably.

The Language-based approach outperforms the standard approach in 100\% of the simulations. 
The LM-based model approach has the following average improvements over the breadth-first baselines: University +582\%, hospital +1004\%, companies +499\%, government +639\%, and \textbf{ALL} +969\%.
\paragraph{Embeddings similarity}
The ability of embeddings to extract context from web application paths and generalize is essential to predict valid directories and URLs.
The results, especially in simulations on general test sets, highlight how the Language model approach successfully uses the context extrapolated from embeddings to achieve significantly better results than the other approaches.

We can observe this by reporting two examples: given two directories, we use the Cosine similarity to measure the top 10 words most similar directories, which should belong to a similar context:
\begin{enumerate} 
\item \textit{article}: ('stories', 0.48), ('academics', 0.43), ('press-release', 0.39), ('press-releases', 0.38), ('video', 0.32), ('authors', 0.32), ('spotlight', 0.32), ('articles', 0.31), ('case', 0.3), ('impact', 0.29)
\item \textit{about}: ('locations', 0.79), ('about-us', 0.75), ('research', 0.75), ('programs', 0.74), ('conditions', 0.7), ('services', 0.68), ('resources', 0.68), ('alumni', 0.68), ('careers', 0.67), ('contact', 0.66)
\end{enumerate}
In both cases, we can see that the words determined similarly by the embeddings represent the same word but slightly different, such as '\texttt{about}' with '\texttt{about-us}' or '\texttt{article}' with '\texttt{articles}'.
In addition, we find other words that relate to the context created by the word under consideration, such as '\texttt{authors}' or '\texttt{stories}' for '\texttt{article}.
This outcome confirms the superiority of LM in generalizing the observed pattern at training time.

\paragraph{Examples of LM Patterns} 
The high average number of successful responses obtained from the LM-based approach testifies to the model's ability to predict valid directories that follow recurring patterns.
For example, let us examine the Language model's predictions on two different URLs:
\begin{enumerate}
\item URL: \textit{/campus-life-events/calendar}. Among the top 10 directories predicted with this URL, we have ['05', '06', '08', '11', 'may', jun'], which refer to days or months of a calendar.
\item URL: \textit{/media}. Among the top directories predicted with this URL, we have ['press-releases', 'news'] that are found in multiple paths in the training dataset and that refer to a similar context.
\end{enumerate}

\section{Related Work}
The emergence of offensive AI in cybersecurity presents a new frontier where artificial intelligence (AI) is leveraged to create sophisticated and automated attacks and enhance the penetration testing process~\cite{kaloudi2020ai, mirsky2023threat}. These attacks represent a new landscape that poses significant challenges and opportunities in cybersecurity, especially with the raising of LLMs and generative AI.

The use of generative AI to enhance directory brute-forcing attacks has yet to be explored. The closest attempt is presented by He et al.~\cite{he2022ai}, where the authors proposed an attack to medical systems by adopting semantic clustering of sentences. 
No much information are reported in terms of data, methodology, and results. 
Similarly, Antonelly et al.~\cite{antonelli2021leveraging} presented an innovative approach using the Universal Sentence Encoder (USE) for semantic analysis. The K-means algorithm and the elbow method were used for clustering to optimize directory brute-forcing (dirbusting), with an improvement in the results of up to 50\% on only eight web applications tested. 

Several other studies have analyzed the threat that offensive AI poses to organizations in other types of attacks.
Bontrager et al.~\cite{bontrager2018deepmasterprints} demonstrated the potential of AI-generated fingerprint deepfakes to compromise biometric systems through dictionary attacks, highlighting the vulnerability of such systems to sophisticated AI techniques. 
Al-Hababi et al.~\cite{al2020man} investigated man-in-the-middle attacks leveraging machine learning to identify services in encrypted network flows.
Li et al.~\cite{li2020feature} presented a generative adversarial network designed to evade PDF malware classifiers, illustrating the ease with which AI can bypass traditional cybersecurity defences. 
Nam et al.~\cite{nam2020recurrent} developed a recurrent GANs-based password cracker aimed at enhancing IoT password security. While intended for defensive purposes, the study also signifies how AI can be repurposed for offensive operations.




\section{Conclusions}
Current directory brute-forcing attacks are notoriously inefficient since they rely on brute-forcing strategies, resulting in an enormous amount of queries for a few successful discoveries.
In this work, we investigated whether the utilization of prior knowledge might result in more efficient attacks. 
We propose two distinct methods that rely on prior knowledge: a probabilistic model and a Language Model-based attack. 
We then experimented with our proposed methodology in a dataset containing more than 1 million URLs, spanning across distinct web app domains such as universities, hospitals, companies, and government. 
Our results show the superiority of the proposed method, with the LM-based approach outperforming brute-force-based approaches in all scenarios (an average performance increase of 969\%). Furthermore, the simple probabilistic approach results effective when the budget of requests is limited (below 100, for stealthier attacks).

The research presented in this paper lays the groundwork for several promising directions for future investigation.
The use of Artificial Intelligence to create sophisticated attacks is a topic that is constantly evolving and growing in cybersecurity, especially with the fast development of Language models.

\paragraph{Advanced Language Models}
Future work could explore improvements of our LM-based architecture, such as attention mechanisms~\cite{vaswani2017attention}, or even Large Language Models~\cite{chang2023survey}.
These models' enhanced understanding of context and semantics could significantly refine the process of predicting web application structures.

\paragraph{Cross-Lingual Contextualization}
Given that directory predictions can be constrained by the language in which a web application is developed, there is potential for leveraging pre-trained embeddings to understand the context better. This understanding could then be transposed to other languages, maintaining the same contextual relevance.

\paragraph{Vulnerability-Specific Language Model}
Additionally, a path that may be explored is the development of a language model trained explicitly on paths and files commonly associated with vulnerabilities. By focusing on these critical areas, the model could help preemptively identify potential security risks, thereby contributing to more proactive cybersecurity measures and showing the feasibility of such attacks.

These areas of future work offer the potential to significantly impact the development of more secure web environments and pose new security challenges to language model usage.

\bibliographystyle{ACM-Reference-Format}
\bibliography{bib}










\end{document}